\documentclass[12pt]{article}
\usepackage{epsfig}

\newcommand{\be}{\begin{eqnarray}}
\newcommand{\ee}{\end{eqnarray}}

\begin{document}

\bibliographystyle{unsrt}
\footskip 1.0cm

\thispagestyle{empty}
\vspace{1in}

\begin{center}{\Large \bf {Nuclear effects in prompt photon production at
the Large Hadron Collider}}\\

\vspace{1in}
{\large  J. Jalilian-Marian$^{1,2}$, K. Orginos$^{1,3}$ and I. Sarcevic$^1$}\\

\vspace{.2in}
{\it $^1$Department of Physics, University of Arizona, Tucson, Arizona
85721\\
$^2$Physics Department, Brookhaven National Laboratory,
Upton NY 11973-5000\\ 
$^3$RIKEN-BNL Research Center,
Brookhaven National Laboratory,
Upton NY 11973-5000\\
}

\end{center}

\vspace*{25mm}

\begin{abstract}

We present a detailed study of prompt photon production cross section
in heavy-ion collisions in the central rapidity region at energy of 
$\sqrt{s}=5.5$ TeV, appropriate to LHC experiment. We include the 
next-to-leading order radiative corrections, $O(\alpha_{em}\alpha_s^2)$, 
nuclear shadowing and the parton energy loss effects. We find that the 
nuclear effects can reduce the invariant cross section for prompt photon 
production by an order of magnitude at $p_t=3$ GeV. 
We discuss theoretical 
uncertainties due to parton energy loss and nuclear shadowing parameters. 
We show that the K-factor, which signifies the importance of 
next-to-leading order corrections, 
is large and has a strong $p_t$ dependence.

\end{abstract}
\newpage

\section{Introduction}

There has been a considerable theoretical and experimental interest in 
studying photon production in heavy-ion collisions at 
BNL's Relativistic Heavy Ion Collider (RHIC) and CERN's Large Hadron 
Collider (LHC) energies \cite{qm99}. Photons produced in 
heavy-ion collisions provide an excellent probe of the properties of the 
dense matter, such as the quark-gluon plasma or the hot hadronic gas,
produced after the collision. Due to the small cross sections of 
electromagnetic interactions, photons can escape the strongly 
interacting matter produced in the collision without further interactions. 

Studying photon production at the RHIC and LHC energies is of 
special interest, as it has been suggested as an elegant signal for detecting 
the formation of a quark gluon plasma (QGP) in heavy-ion collisions 
\cite{mt}. However, photons can be produced at different
stages of the heavy ion collision and thus have different origin.  
For example, photons can be produced at the 
early stages of the collision through QCD processes such as 
$qg \rightarrow q\gamma$ or they can be emitted from a thermalized quark
gluon plasma or hadronic gas. Another source of produced photons is 
decay of hadrons such as pions and etas produced in the heavy ion 
collision \cite{kls}. Furthermore, different processes give the dominant 
contribution at different $p_t$'s. Therefore, it is quite difficult to 
make reliable predictions for the absolute number of 
photons produced in a heavy 
ion collision \cite{hdsg,sri}. 

Prompt photons are an important background to thermal photons 
in the low to intermediate 
$p_t$ region and are dominant in the high $p_t$ region. Therefore, it is 
extremely important to be able to calculate their production cross section 
reliably. Fortunately, one can use perturbative QCD in the high $p_t$ 
region to calculate the prompt photon production cross section. In this work 
we continue our study of the production of high $p_t$ ($p_t > 3$ GeV) 
prompt photons in heavy ion collisions\cite{jos1}. Prompt photons are 
produced either directly in the hard 
collisions of the partons inside the nuclei like $qg \rightarrow q\gamma$ 
or through bremsstrahlung of quarks and gluons produced in the hard collision 
such as $qg \rightarrow qg\gamma$. We include all next-to-leading order,
$O(\alpha_{em}\alpha_s^2)$ QCD processes \cite{abfs} as well as nuclear 
shadowing and medium induced parton energy loss effects. We discuss and 
estimate the theoretical uncertainties  
due to different choices of nuclear shadowing and energy loss 
parameters. We show that next-to-leading (NLO) corrections are large and 
must be included to make reliable predictions. 

In section I we review the prompt photon production in hadronic
collisions in next-to-leading order. Based on QCD factorization theorems, 
we write down the expression for prompt photon production and list
the hard partonic processes involved. In section II we discuss the 
nuclear effects such as shadowing and energy loss involved in production 
of prompt photons in heavy ion collisions. In section III we present our
results for the prompt photon production invariant cross section, 
$E{d\sigma \over d^3p}$, at LHC energies and an estimate of the  
theoretical uncertainties due to variation of nuclear shadowing and energy
loss parameters. We show that the effective K-factor, defined as the ratio
of NLO to LO cross sections in heavy ion collisions is large and has a 
strong $p_t$ dependence. This clearly shows the importance of including
next-to-leading order contributions to prompt photon production.  We conclude 
with a discussion of prospects for detecting nuclear effects by measuring 
prompt photons at LHC.  

\section{Prompt photon production in pQCD}

Using factorization theorems and  perturbative QCD, the inclusive cross section
for prompt photon production in a hadronic collision can be written  
as a convolution of parton densities in a hadron with the hard
scattering cross section and the parton to photon fragmentation function:  

\be
E_{\gamma} \frac{d^3\sigma}{d^3p_\gamma}(\sqrt s,p_\gamma) 
=\int dx_{a}\int dx_{b} \int dz \sum_{i,j}^{partons}F_{i}(x_{a},Q^{2}) 
F_{j}(x_{b},Q^{2}) D_{c/\gamma}(z,Q^2_f)
E_\gamma {d^3\hat{\sigma}_{ij\rightarrow c X}\over
d^3p_\gamma}
\label{eq:fact}
\ee
where the $F_{i}(x,Q^{2})$ is the i-th parton distribution in a nucleon,
$x_a$ and $x_b$ are the fractional momenta of incoming partons, 
$ D_{c/\gamma}(z,Q^2_f)$ is the photon fragmentation function
with $z$ being the fraction of parton energy carried by the photon. The
parton-parton cross sections, 
${d^3\hat{\sigma}_{ij\rightarrow c X}\over d^3p_\gamma}$, include all
processes up to and including $O(\alpha_{em}\alpha_s^2)$, such as leading-order 
subprocesses: 
\be
q + \bar q &\rightarrow & \gamma + g \nonumber\\
q +  g &\rightarrow & \gamma + q
\label{eq:lo}
\ee 
and the next-to-leading order subprocesses, 
\be 
q + q &\rightarrow &  q + q + \gamma \nonumber\\
q + \bar q &\rightarrow & q + \bar q + \gamma \nonumber\\
q + q' &\rightarrow & q + q' + \gamma \nonumber \\
q + \bar q &\rightarrow & q' + \bar q' + \gamma \nonumber\\
q + \bar q' &\rightarrow & q + \bar q' + \gamma
\label{eq:nlo}
\ee
We refer the reader to \cite{abfs} for a complete list of all 
$O(\alpha_{em}\alpha_s^2)$ processes.  There are two classes of 
subprocesses, ``direct'' production, which does not 
have convolution with the 
fragmentation function and the ``bremsstrahlung'' contribution 
that has 
convolution with the fragmentation function.  Direct subprocesses 
contribute to leading order as well as next-to-leading order, while 
bremsstrahlung processes only contribute at the next-to-leading 
order.  Clearly, only the sum of these two contributions constitutes 
the full 
NLO calculation of the direct photon production.  
It is important to note that 
in order 
to include all $O(\alpha_{em}\alpha_s^2)$ processes, it is necessary to 
include processes which formally look order $O(\alpha_{em}\alpha_s^3)$. This
is because some of the terms in (\ref{eq:nlo}) have a divergence
proportional to $1/\alpha_s$ which makes those processes  
$O(\alpha_{em}\alpha_s)$. This divergence comes from the quark and emitted 
photon being collinear. The reader is referred to \cite{acfgp} for a 
discussion of the collinear divergences.

The nucleon structure functions, $F_{i}(x,Q^2)$, and parton to
photon fragmentation functions, $D_{c/\gamma}(z,Q^2_f)$, are also 
evaluated at the next-to-leading order. We use the MRS$99$ set 
for nucleon structure functions \cite{mrs99} and Bourhis et al. 
parameterization of the photon fragmentation functions 
\cite{bfg}.  
Structure functions, fragmentation functions and the running coupling constant 
depend on the 
factorization, fragmentation and renormalization scales respectively  
which are usually taken to be the same and 
proportional to the photon transverse momentum $p_t$. 
Aurenche et al. \cite{aur} have studied the dependence of the prompt 
photon cross section 
on the choice of scale and have shown that the choice of $Q= p_t/2$ gives
a very good description of prompt photon production in hadronic collisions.
Therefore, we will use $Q=p_t/2$ in our calculation.
The running coupling  constant $\alpha_{s}(Q^{2})$, calculated to 
next-to-leading order, is given by  
\be
\alpha_s (Q^2)={12\pi\over (33-2N_f)\ln Q^2/\Lambda^2}
\bigg[1-{6 (153-19N_f)\ln\ln Q^2/\Lambda^2 \over 
(33-2N_f)^2\ln Q^2/\Lambda^2}\bigg] 
\ee
where $Q^2$ is the renormalization scale, $\Lambda$ is the QCD scale 
parameter and $N_f$ is the number of flavors.

\begin{figure}[htp]
\centerline{
    \epsfxsize 3.0 truein \epsfbox{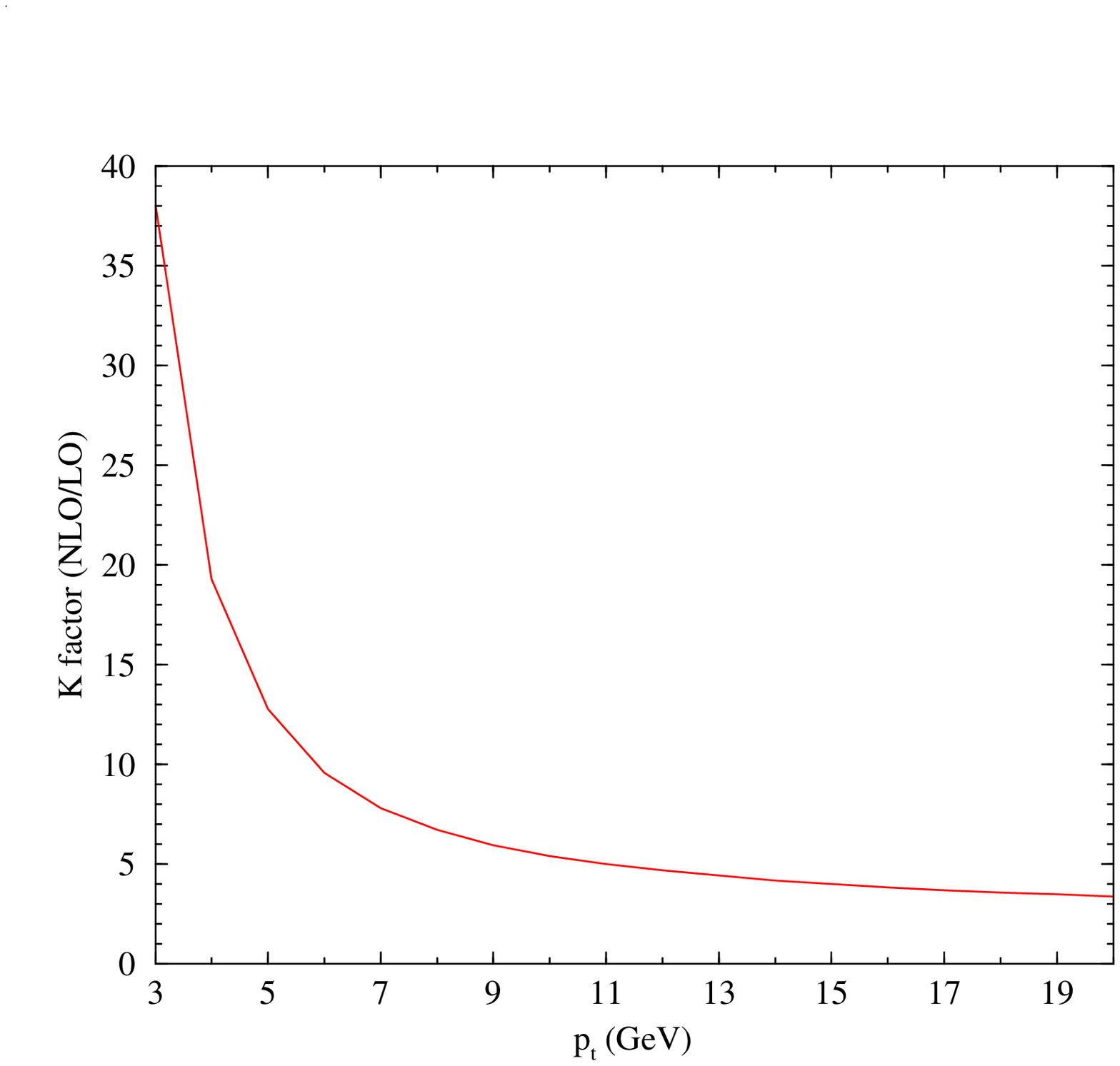}
    \hskip 0.1 truein
    \epsfxsize 3.0 truein \epsfbox{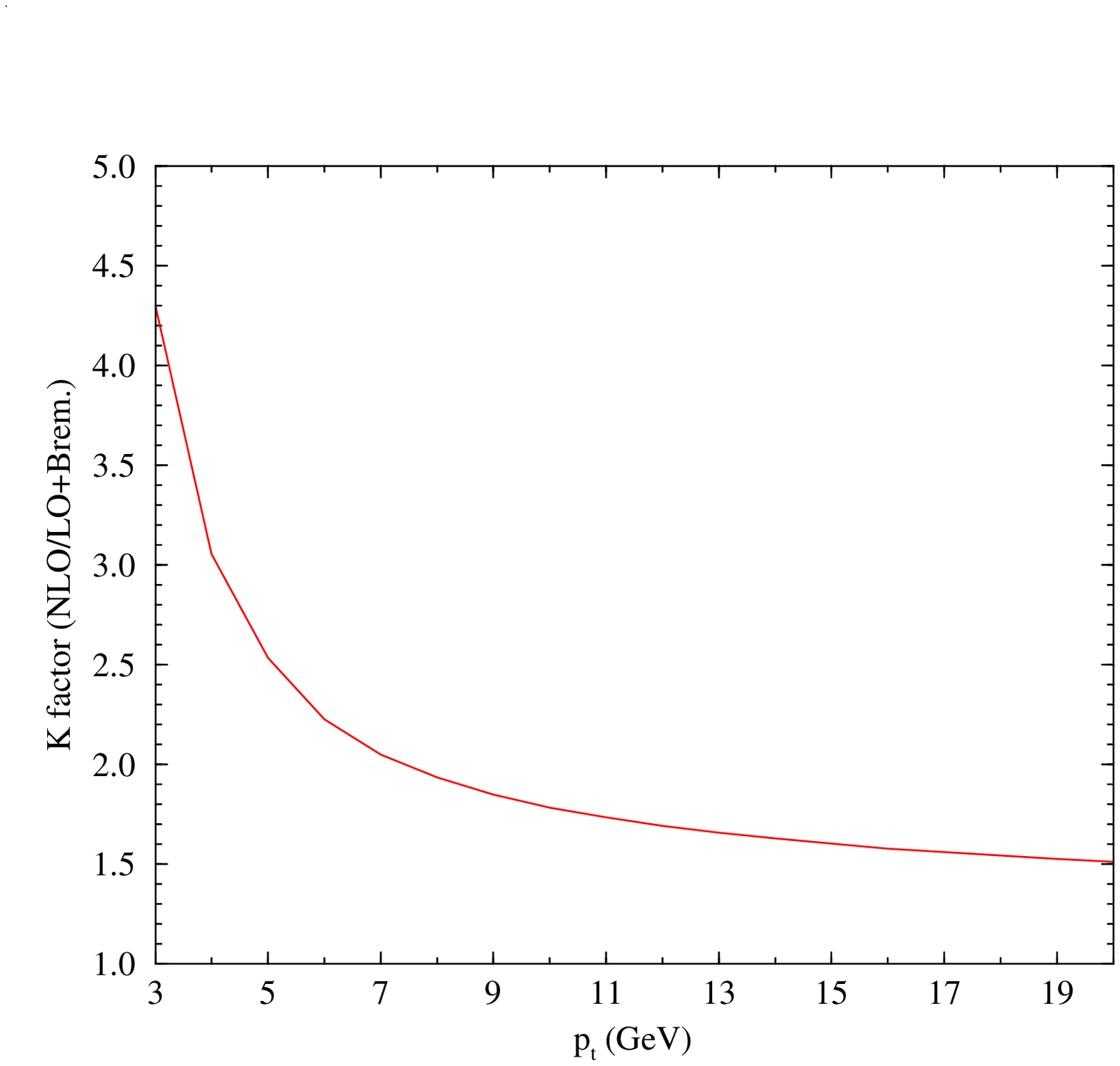}
 }
\caption{a) The hadronic K-factor defined as the ratio of NLO 
prompt photon cross section, $E{d\sigma \over d^3p}$ to 
the LO cross section,  
and b) the K-factor defined as the ratio of NLO to the LO plus 
bremsstrahlung cross sections.}
\label{fig:kflo}
\end{figure}

In Figure (\ref{fig:kflo}a) we illustrate the importance of including the 
next-to-leading order contributions by calculating the 
ratio of NLO to LO cross sections, the so-called 
K-factor.  Clearly, the NLO corrections are large and $p_t$ dependent.  
It would be useful to go even beyond the NLO to make sure that 
the higher order corrections are not even larger.   
It is, however, clear that 
a leading order calculation in the LHC kinematic region is
meaningless since the next-to-leading order corrections are huge.

In Figure (\ref{fig:kflo}b) we show an alternative definition of the 
K-factor defined as $K\equiv NLO/(LO+Brems.)$ sometimes used in the 
literature \cite{hdsg}. The main reason for this definition was 
inclusion of an incomplete set of Bremsstrahlung diagrams in previous 
works on prompt photon cross sections \cite{hdsg} 
and we show it here for comparison.  

In principle, in the region 
where $x_t \sim p_t/\sqrt{s}$ is large, 
i.e. in the very low and very high $p_t$ region of phase space, 
one would need to resum 
certain logarithmic terms, of the form $\alpha_s \ln (1-x_t)$ and 
$\alpha_s \ln x_t$.  
However, 
these 
resummations are not very important for the range of $p_t$ and $\sqrt{s}$
that are considered in this work and will be neglected \cite{wv}.

There is also an alternative approach to calculating prompt photon
cross sections which uses the LO cross sections as well as LO structure
and fragmentation functions. In order to reproduce the experimental
data in this approach, it is necessary to include a phenomenological 
parameter which is loosely identified as the ``intrinsic'' momentum 
of the initial state partons \cite{kt}. This intrinsic momentum is 
generated 
by the initial state radiation of quarks and gluons. There is however
no theoretical calculation of these effects at the moment and one
has to model them by generalizing the standard definitions
of the parton distribution functions from $q(x,Q^2)$ and $G(x,Q^2)$ to
$q(x,k_t^2,Q^2)$ and $G(x,k_t^2,Q^2)$. One then introduces a weight
function, typically a Gaussian, with a width $<k_t^2>$ which represents
the intrinsic transverse momentum of the initial state partons.
Assuming some reasonable range in $k_t$, these intrinsic momenta 
are integrated over in the final result. This approach gives 
fairly good description of most of the experimental results.

It is important to realize that this intrinsic momentum $k_t$ 
grows with energy and can be as large as $\sim 1-2$ GeV in fixed
target experiments and $\sim 5$ GeV in the Tevatron \cite{rep}. Therefore, 
strictly speaking, it is not an intrinsic momentum. In addition,  
the value of $k_t$ is 
process dependent and thus can not be taken as 
universal parameter.  It might be possible to extract its value 
from the data in different experiments, but its value can not be
predicted. In our work, we will not follow this approach but instead we will 
use NLO perturbative QCD formalism because it seems theoretically
more rigorous and self-consistent. We note that 
NLO calculations give fairly good
agreement with the experimental data for 
$p_t > 3.5$GeV at lower energies ($\sqrt s \leq 63$ GeV) \cite{aur} and for 
$p_t > 10$GeV at higer energies ($\sqrt s = 1800$ GeV) \cite{rep}. However, 
it is worth pointing out that 
there are claims of 
inconsistencies among various data sets from different experiments
at low energies \cite{aur,rep}. 
In addition, one should keep in mind that 
our study extends down to lower values of $p_t$, where NLO 
predictions may not be reliable.  

\section{Prompt photon production in heavy ion collisions}

To calculate the prompt photon production cross section in heavy
ion collisions, we will use eq. (\ref{eq:fact}) modified for nuclear
effects:

\be
E_{\gamma} \frac{d^3\sigma^{AB}}{d^3p_\gamma}(\sqrt s,p_t) 
=\int dx_{a}\int dx_{b} \int dz \sum_{i,j}^{partons}F_{i}^A(x_{a},Q^{2}) 
F_{j}^B(x_{b},Q^{2}) D_{c/\gamma}(z,Q^2_f)
E_\gamma {d^3\hat{\sigma}_{ij\rightarrow c X}\over
d^3p_\gamma}
\label{eq:nfact}
\ee
where the $F_{i}^A(x,Q^{2})$ is the i-th parton distribution in a nucleus
and $ D_{c/\gamma}(z,Q^2_f)$ is the photon fragmentation function
in a nuclear environment. The partonic processes (a partial list) are given by 
(\ref{eq:lo}, \ref{eq:nlo}) as before. The nuclear structure function 
$F_{i}^A(x,Q^{2})$ and fragmentation function  $ D_{c/\gamma}(z,Q^2_f)$ 
are not known to next-to-leading order, but rather they are NLO nucleon 
structure and fragmentation functions modified for nuclear effects.
Strictly speaking, therefore, our calculation of prompt photon production
in heavy ion collisions is not a next-to-leading order calculation but
is the most complete calculation performed so far. Below, we will describe
the nuclear modifications to structure functions and fragmentation 
functions in detail.

\subsection{The Nuclear Shadowing Effect}

Calculation of the prompt photons production cross section in nuclear 
collisions requires knowledge of the nuclear structure functions 
$F_{i}^A (x,Q^2)$, usually measured in Deep Inelastic Scattering (DIS) 
of electrons on nuclei. It is an experimental fact that 
$F_{i}^A (x,Q^2)\neq A F_{i}^N (x,Q^2)$ where  $F_{i}^N (x,Q^2)$ is the
free nucleon structure function. This modification has a strong $x$ 
dependence and is due to having different nuclear effects in different 
region of phase space.  
At small values of $x$, for instance $x < \sim 0.07$, the nuclear
structure function is less than nucleon structure function  scaled by $A$.
This is known as shadowing. As $x$ grows bigger, nuclear structure
functions get bigger than the free nucleon structure function. This is 
known as anti-shadowing. As $x$ further increases, nuclear structure 
functions become less than the free nucleon ones again which is known as the 
EMC effect. In this section, we will concentrate on shadowing and 
anti-shadowing regions since those are the kinematic regions where the 
prompt photon production takes place. We refer the reader to \cite{arn}
for a recent review of nuclear shadowing.

An intuitive explanation of nuclear shadowing and anti-shadowing effects 
in DIS depends 
on the reference frame of the nucleus. Of course, the physical observables 
such as structure functions can not depend on our choice of reference frame. 
However, working in different frames helps one identify the different
physical mechanisms involved. In the rest frame of the nucleus and 
in perturbative QCD,  shadowing of nuclear structure functions in DIS 
is due to multiple interaction of the $q \bar q$ component of the photon
wave function (emitted by the electron) with the nucleus. The amplitude 
of $q \bar q A$ interaction is mostly 
imaginary at small $x$ and multiple interactions of the pair with the nucleus
introduces a phase difference between different amplitudes which leads
to a destructive interference. This in turn reduces the nuclear cross section. 
In a non-perturbative description of shadowing, the photon is resolved in 
terms of its hadronic fluctuations which in turn multiply interact with
the nucleus. The multiple interactions again reduce the nuclear cross
sections due to destructive interference. In this frame, anti-shadowing
at larger $x$ is due to a large real part of the interaction amplitudes
which interfere constructively with the imaginary part and lead to an
enhancement of the nuclear cross section (structure functions).

In the infinite momentum frame where the nucleus is moving very fast,
shadowing is caused by high parton density effects small $x$. The
small $x$ partons have a large longitudinal wavelength and can spatially
overlap and recombine. These recombination effects reduce the nuclear parton
number densities and hence the nuclear cross sections. Working in this
frame enables one to treat nuclear shadowing and parton saturation
in nucleons on the same footing due to the identical physical mechanism
involved in both. Anti-shadowing is due to longitudinal momentum 
conservation (momentum sum rule) in this frame.

Even though there has been considerable amount of theoretical work
done on nuclear shadowing and impressive progress made in understanding
the physical principles of nuclear shadowing \cite{arn}, we are far from 
having a precise and quantitative description of nuclear shadowing.
In practice, one measures the nuclear structure functions in deep inelastic
scattering of leptons on nuclei \cite{nmc}. The measured structure functions
are then used in nuclear collisions. The scale dependence of the nuclear
structure functions is even less understood due to the limited range of
$Q^2$ covered in fixed target experiments. Also, shadowing of gluons
is not well understood due to the fact that they can not be directly 
measured in DIS experiments. The working assumption is that high
parton density effects are negligible and DGLAP evolution equations
are valid in which case the gluon distribution function can obtained
from the scaling violation of the $F_2$ structure functions. This 
assumption, however, will break down at small values of $x$ due to
high parton density effects \cite{mnmob} and one will need to measure 
the gluon distribution function differently.

In this work we will use two different parameterizations of nuclear 
structure functions due to Benesh, Qiu, Vary \cite{bqv} and Eskola et al.
\cite{eks98}. Both parameterizations fit the current experimental
data quite well even though they are somewhat different. Also, since there
are no experimental data at the small $x$, high $Q^2$ ($Q^2 > 1GeV^2)$ region, 
any parameterization of nuclear structure functions in this region is 
subject to large uncertainties. This is somewhat important for RHIC but
becomes crucial for LHC. An $eA$ collider such as the proposed eRHIC
is urgently needed and would greatly improve our knowledge of nuclear
structure functions in the small $x$, high $Q^2$ region as well as reducing
the theoretical uncertainties in the larger $x$ region.

\begin{figure}[htp]
\centering
\setlength{\epsfxsize=10cm}
\centerline{\epsffile{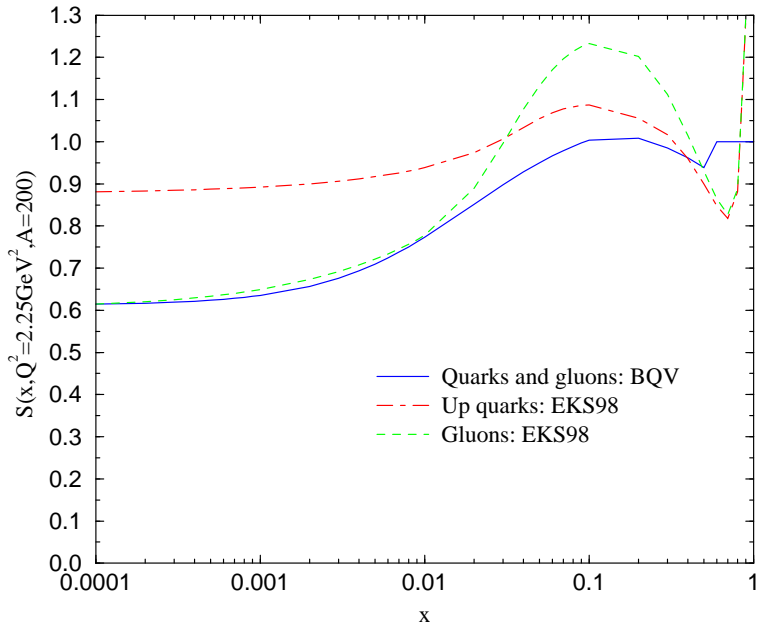}}
\caption{The nuclear shadowing ratio as parameterized by BQV \cite{bqv} 
 and
EKS98 \cite{eks98}).}
\label{fig:eksbqv}
\end{figure}

The parametrization of the nuclear shadowing function proposed by 
Benesh, Qiu and Vary is given by \cite{bqv} 
\be
S(x,A)=\left\{\begin{array}{ll}
\alpha_3 -\alpha_4 x & x_0 <x\leq 0.6 \\
(\alpha_3 -\alpha_4 x_0)\frac{1+k_q \alpha_2 ({1/x}-1/x_{0})} 
{1+k_q A^{\alpha_1}({1/x}-1/x_{0})} 
&x\leq x_{0}\\
\end{array}
\right \}
\ee
It gives a good description of all EMC, NMC and E665 data \cite{nmc}.  
The parameters $k_q$, $\alpha_{1}$, $\alpha_2$, $\alpha_3$ and $x_0$ 
are fitted to deep inelastic data  for the ratio $F_2^A(x,Q^2)/F_2^D(x,Q^2)$ 
and can be found in \cite{bqv}. In this parametrization, nuclear
structure functions are independent of $Q^2$ and shadowing of gluons is
assumed to be the same as quarks. A more recent parameterization of the 
nuclear structure function is that of Eskola et al. \cite{eks98} 
that 
also fits the existing experimental data quite well, has $Q^2$
dependence and distinguishes between quark and gluon 
structure functions \cite{eks98}. 
We show the nuclear shadowing ratio defined as $S\equiv F_2^A/AF_2^N$ 
in Figure (\ref{fig:eksbqv}) using the parameterizations of BQV \cite{bqv}
and EKS \cite{eks98}. Clearly, the two are quite different although both
fit the experimental data fairly well. This signifies the need for
a high energy collider experiment such as eRHIC to improve on the
current measurements of the nuclear structure functions. 

In this work, we will use both parameterizations
and investigate the dependence of prompt photon production cross sections
on our choice of nuclear structure functions.

\subsection{Medium induced parton energy loss effects}

Another difference between hadronic and heavy ion collisions is in the
multiplicity of the final state particles produced. In high energy heavy ion
collisions and in central rapidities, the multiplicity of particles
produced per unit rapidity is much larger than that in hadronic 
collisions. Therefore, many body effects such as secondary collisions,
which are negligible in hadronic collisions become important 
in high energy heavy ion collisions. Another example is the medium induced
energy loss. In hadronic collisions, since the number density of particles
per rapidity produced is small, one can neglect the multiple interactions
of the produced partons with each other. On the other hand, in high 
energy heavy ion collisions, due to high multiplicities per unit rapidity, 
the multiple interactions of produced partons with each other can not
be neglected. As a consequence of multiple interactions with the medium, 
produced particles can lose energy before hadronizing. This affects 
their energy and momentum spectrum. 
Energy loss of energetic partons passing through a dense medium has
been a hot topic lately \cite{eloss}. There has been a considerable
progress made in understanding the different forms of the energy 
loss in different limits. It has been shown, for a finite size
medium, that parton energy loss increases with increasing parton 
energy \cite{eloss}.  
Current calculations of energy loss effects are done at the
leading order $O(\alpha_s)$ and a next-to-leading order calculation
is not presently available.  

A rigorous treatment of energy loss effects in 
a heavy ion collision is extremely complicated and beyond the 
scope of this work. Rather, we will take a phenomenological approach 
to medium induced energy loss in nuclear collisions and use a model
of Wang, Huang and Sarcevic \cite{hsw},
to estimate energy loss effects in high energy heavy ion
collisions. In this commonly used model, it is assumed that the
main effect of multiple interactions in the medium can be accommodated 
by modifying 
the photon fragmentation functions. In the central rapidity
region, parton produced in the hard collision is traversing the
nuclear medium and losing its energy as a result of multiple interactions
with the deconfined medium. This parton will hadronize outside the 
nuclear medium but with a reduced 
energy.

In the energy loss model of Wang, Huang and Sarcevic \cite{hsw}, 
the parton to photon fragmentation function, $zD^0_{a/\gamma}(z,Q^2_f)$, 
which gives 
the probability for a parton to fragment into a photon, is modified 
to include multiple scatterings of the fragmenting parton from the 
nuclear medium before it fragments. The nuclear fragmentation function 
$zD_{a/\gamma}(z,Q^2_f)$ is given in terms of the photon fragmentation 
function $zD^0_{a/\gamma}(z,Q^2_f)$ by \cite{hsw} 
\be
zD_{a/\gamma}(z,\Delta L,Q^2_f)=
\frac{1}{C^a_N}\sum_{n=0}^N P_a(n) \bigg[z^a_nD^0_{a/\gamma}(z^a_n,Q^2_f)
+ \sum_{j=1}^{n} \bar{z}^j_aD^0_{g/\gamma}(\bar{z}^j_a,Q^2_f)\bigg]
\label{eq:frag}
\ee
where $z^a_n=z/(1-(\sum_{i=0}^{n}\epsilon^a_i)/E_t)$, 
$\bar{z}_j^a=zE_t/\epsilon_j^a$ and $ P_a(n)$ is the probability that
a parton of flavor $a$ traveling a distance $\Delta L$ in the nuclear
medium will scatter $n$ times. It is given by

\be
P_a(n) = \frac{(\Delta L/\lambda_a)^n}{n!} e^{-\Delta L/\lambda_a}, 
\ee
and 
$C^a_N=\sum_{n=0}^N P_a(n)$.  

The first term in Eq. (\ref{eq:frag}) corresponds to the fragmentation of 
the leading parton $a$ with reduced energy $E_t- \sum_{i=0}^{n}\epsilon^i_a$ 
after $n$ gluon emissions and the second term comes from the $j$-th 
emitted gluon having energy $\epsilon^j_a$, where $\epsilon^j_a$ is the
energy loss of the parton a after j-th scattering. Since we are studying
the energy loss effects only phenomenologically, we will consider two
different cases for the energy loss per collision, $\epsilon^j_a$. First
we will take it to be constant, as considered for example, in \cite{xnw}.
We will also consider energy dependent energy loss in the form of 
$\epsilon^j_a = \alpha_s \sqrt{\mu^2\lambda_a E^j_a}$, where $E^j_a$ is 
the energy of the parton $a$ after $j$ scatterings, 
$\lambda_a$ is the inelastic mean free path of parton $a$ and $\mu^2$ 
represent a screening mass generated by the plasma and serves as an 
infrared cut off. It should be emphasized that the general form of the
energy loss per scattering, $\epsilon^j_a$, is theoretically unknown 
and we consider 
two possible cases. In order to study 
theoretical uncertainties involved, we will take several values for the 
constant
energy loss and in 
the case of energy-dependent energy loss we will 
vary parameters $\mu^2$ and $\lambda_a$.  

\begin{figure}[htp]
\centerline{
    \epsfxsize 3.0 truein \epsfbox{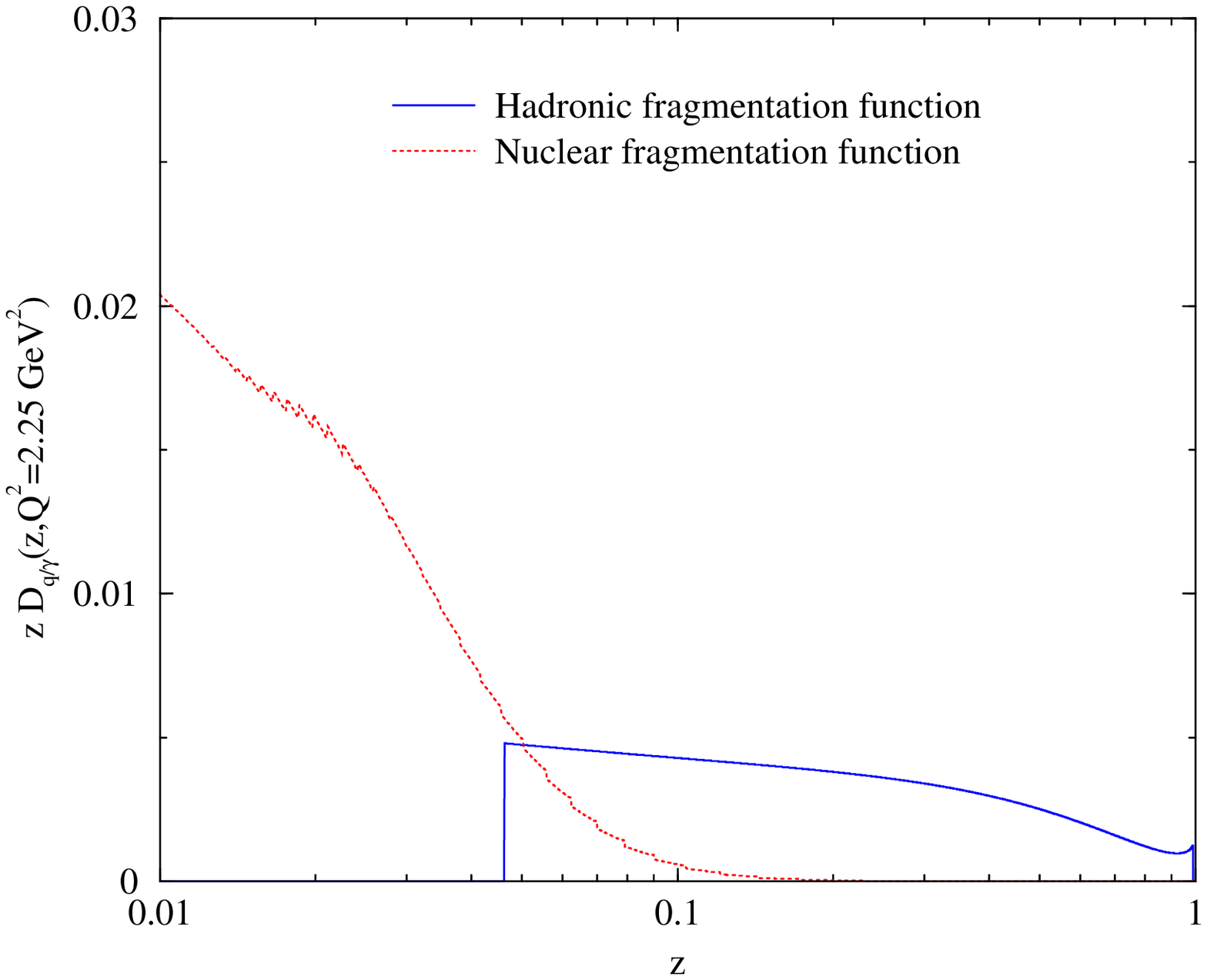}
    \hskip 0.1 truein
    \epsfxsize 3.0 truein \epsfbox{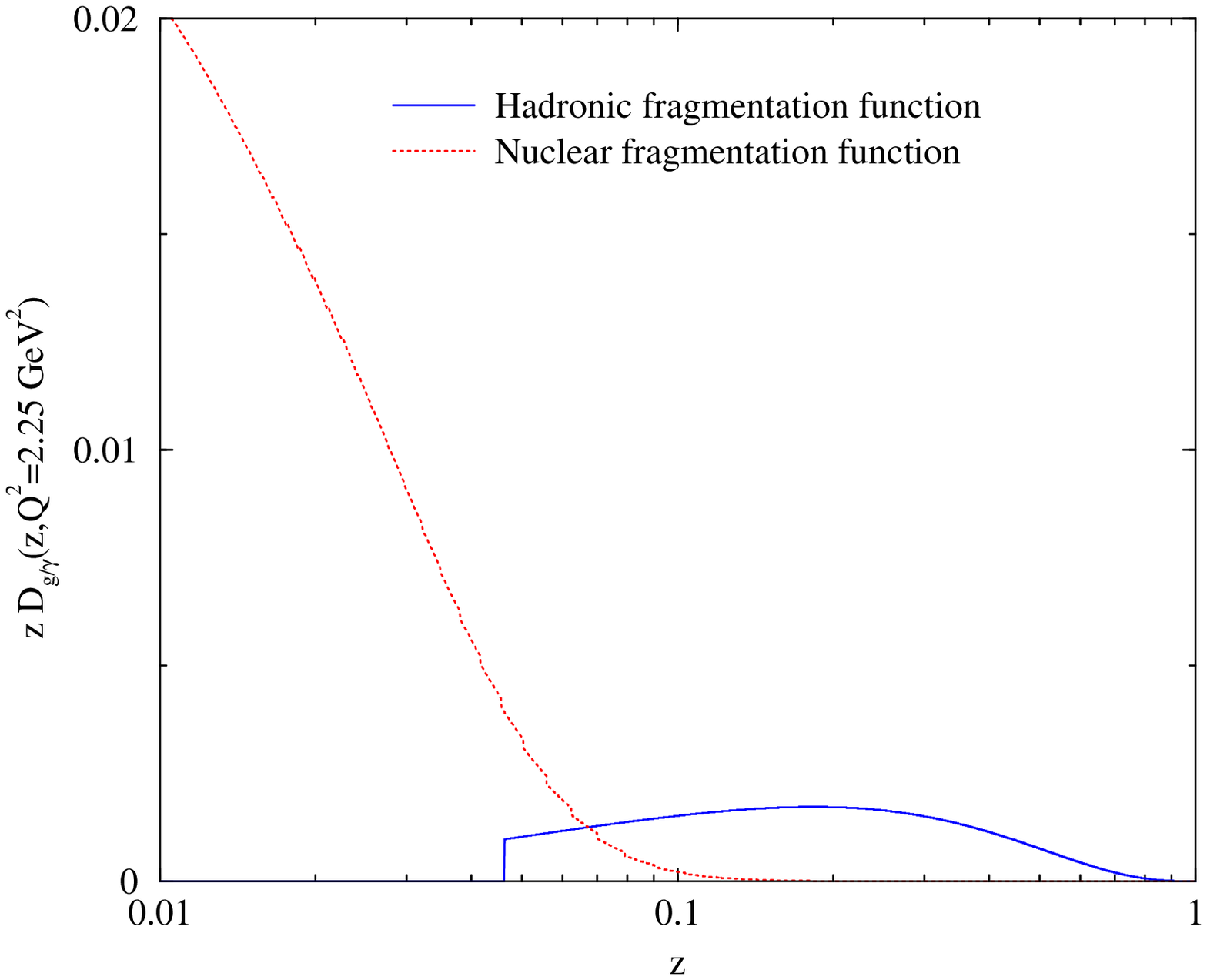}
}
\caption{The photon fragmentation function, 
a) $zD^0_{q/\gamma}(z,Q^2_f)$ 
(solid line) \cite{bfg} and the nuclear 
fragmentation function 
$zD_{q/\gamma}(z,Q^2_f)$ obtained using (\ref{eq:frag}) (dashed line) and b) 
$zD^0_{g/\gamma}(z,Q^2_f)$ (solid line) \cite{bfg} and the corresponding 
nuclear 
fragmentation function 
$zD_{g/\gamma}(z,Q^2_f)$ (dashed line).}
\label{fig:fraglhc}
\end{figure}

The nuclear fragmentation
functions of quarks and gluons as defined in (\ref{eq:frag}), with
energy dependent energy loss, are shown in 
Figure (\ref{fig:fraglhc}).  Here we have taken 
$\mu^2=1$GeV$^2$ and $\lambda_a=1$fm. The constant energy loss case 
is quite similar. For the photon fragmentation function, 
$zD^0_{a/\gamma}(z,Q^2_f)$, we use the parameterization of \cite{bfg}.
The parton to photon fragmentation functions in a nuclear medium are 
enhanced at small $z$ and suppressed at large $z$ as compared to
the fragmentation functions in vacuum. This is due to the fact that
high energy (high $z$) partons multiply scatter from the medium and
lose their energy which shifts their energy fraction $z$ to a smaller
value.  The nuclear fragmentation function obtained assuming constant 
energy loss per scattering has qualitatively the same behavior as 
a function of $z$.

As one goes to higher energies in heavy ion collisions, one probes
smaller and smaller energy fractions $z$ in the fragmentation functions.
The fragmentation functions, 
$zD^0_{q/\gamma}(z,Q^2_f)$, 
are fit to the experimental data
and parameterized. However, the existing data does not cover the
$z$ ranges which will be explored by LHC. Therefore, the current
parameterizations of parton to photon fragmentation functions are set
equal to zero below some energy fraction,  $z < z_0 \sim 0.01$.  This is 
shown in Figure (\ref{fig:fraglhc}).  In the
kinematic region appropriate to LHC, however, on will need to know
the fragmentation functions below this energy fraction. Therefore, we
use both the standard (hadronic) fragmentation functions which are
zero below $z_0$ and another parameterization which is identical to
the standard one for $z > z_0$ but is set equal to a constant for $z < z_0$, 
i.e. 
$zD^0_{q,g/\gamma}(z,Q^2_f)= 
zD^0_{q,g/\gamma}(z=z_0,Q^2_f)$ for $z \leq z_0$.  
We have checked that the difference 
between the two parameterizations leads to a negligible ($< 1\%$) 
difference in the nuclear cross section. The reason is that the 
average $z$ is sufficiently large ($ z\sim 0.1$) and thus the results are 
not sensitive
to the variation of fragmentation functions in the small $z$ region.
Also, partonic cross sections at small $z$ are power suppressed and do 
not contribute significantly.

\begin{figure}[htp]
\centerline{
        \epsfxsize 3.0 truein \epsfbox{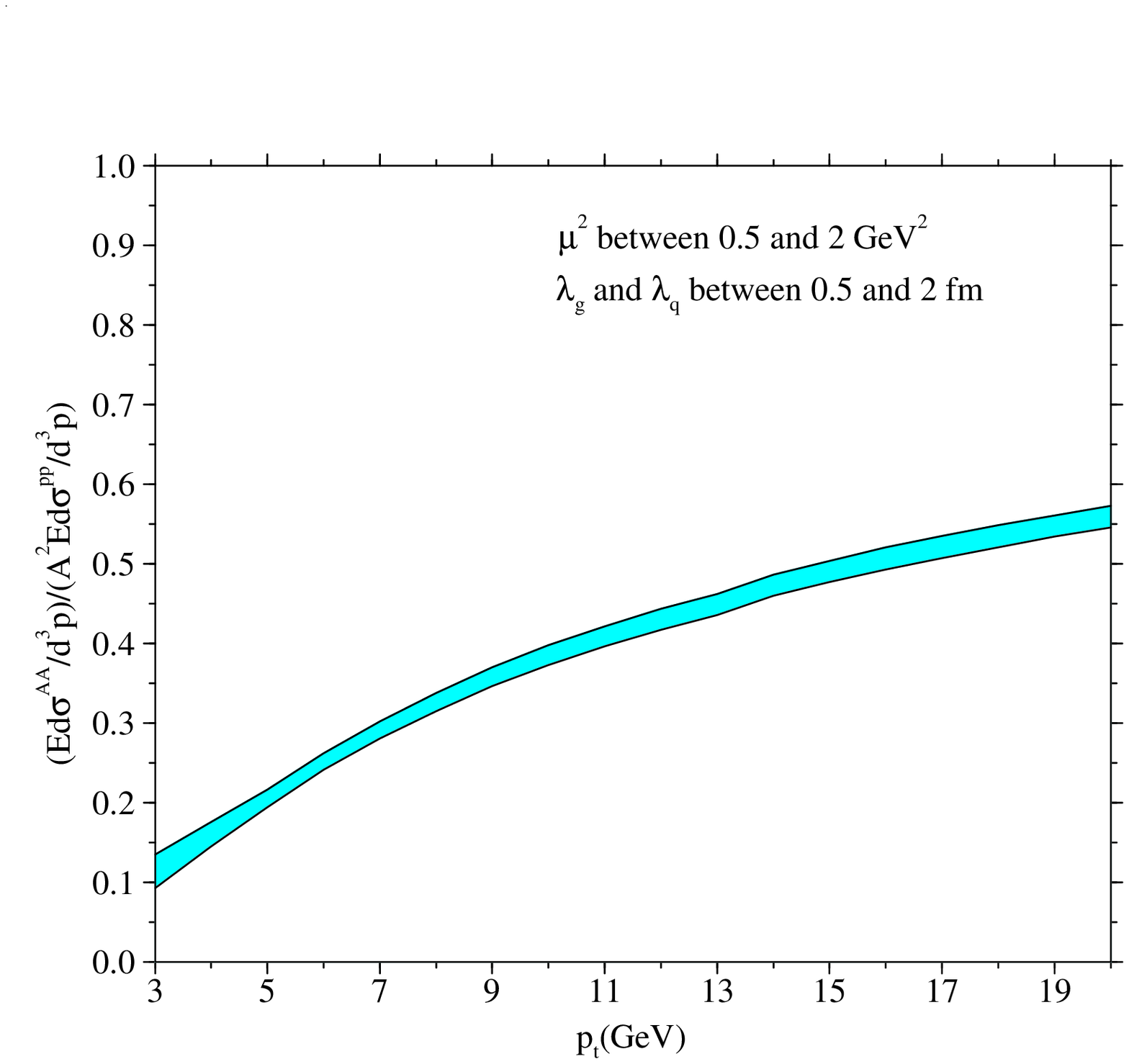}
}
\caption {The uncertainty in prompt photon cross section 
at $\sqrt{s}=5.5$ TeV due to variation of energy-dependent energy loss 
parameters.} 
\label{fig:newlhc21diffmulvspt}
\end{figure}

The parton mean free path $\lambda_a$ and the screening mass $\mu^2$ 
are largely unknown
and in principle depend on the parton species and the medium temperature.
In this work, we treat $\mu^2$ and $\lambda_a$ as unknown parameters 
and show the dependence of our results on a physically reasonable 
variation of them in Figure (\ref{fig:newlhc21diffmulvspt}).  
By varying these parameters, $\mu^2$ and $\lambda_a$, we study 
theoretical uncertainties.  Experimental determination of these 
parameters would be difficult.  
Energy dependent form of energy loss, 
$\epsilon^j_a = \alpha_s \sqrt{\mu^2\lambda_a E^j_a}$, 
is, strictly speaking, 
valid only for coherent photon radiation.  Here, we have considered 
this form 
just as a phenomenological expression that gives photon $p_t$ 
distribution that we could 
compare with the constant
energy loss case.  The precise form of energy loss in a realistic
nuclear collision, such as those at RHIC and LHC energies, 
is not well understood at the moment and is expected to be 
extremely complicated.

\section{Discussion}

\begin{figure}[htp]
\centering
\setlength{\epsfxsize=10cm}
\centerline{\epsffile{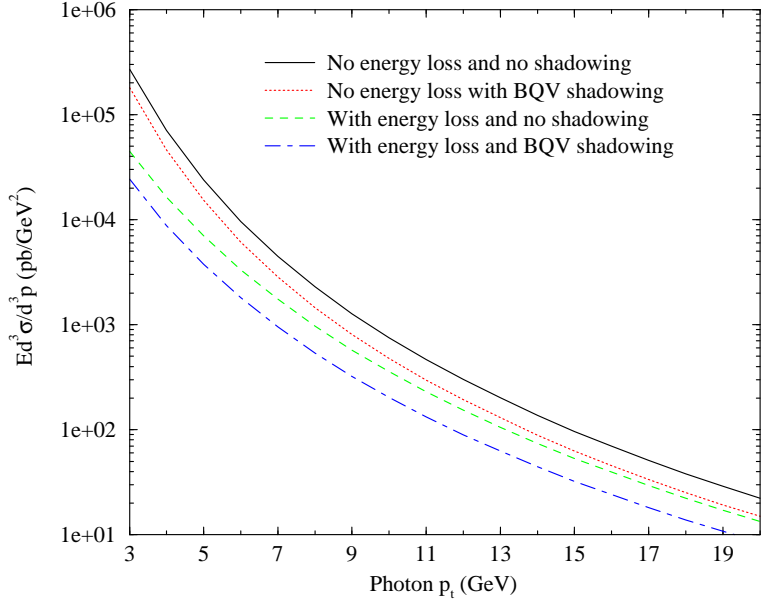}}
\caption{Prompt photon cross section in the central rapidity region at 
$\sqrt{s}=5.5$ TeV without nuclear effects (solid line), 
with BQV \cite{bqv}  
shadowing 
and no parton energy loss (dotted line), with 
parton energy loss 
and no shadowing (dashed line) and 
with BQV \cite{bqv}  
shadowing  
and parton energy loss (dashed-dotted 
line).  
Parton energy loss is
taken to be energy dependent,
$\epsilon^j_a = \alpha_s \sqrt{\mu^2\lambda_a E^j_a}$,
with
$\mu^2=1$GeV$^2$ and $\lambda_a=1$fm.
}
\label{fig:newlhcbqvmu1l1}
\end{figure}

\begin{figure}[htp]
\centering
\setlength{\epsfxsize=10cm}
\centerline{\epsffile{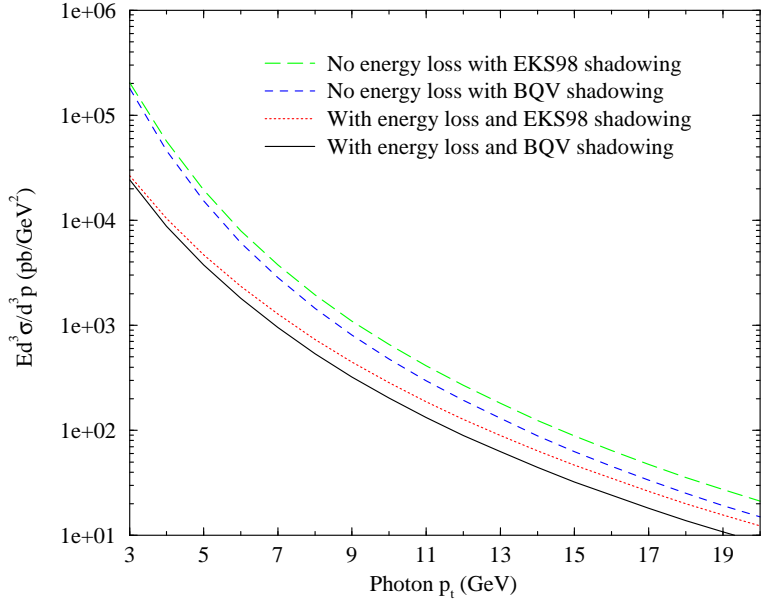}}
\caption{Prompt photon cross section in the central rapidity region 
at $\sqrt{s}=5.5$ TeV obtained with EKS98 \cite{eks98} 
shadowing and without parton 
energy loss (dotted line), with 
parton energy loss (dashed line), 
with BQV shadowing \cite{bqv}
 and no parton energy loss (solid line) and with energy loss 
(dot-dashed 
line).
Parton energy loss is
taken to be energy dependent,
$\epsilon^j_a = \alpha_s \sqrt{\mu^2\lambda_a E^j_a}$,
with
$\mu^2=1$GeV$^2$ and $\lambda_a=1$fm.
}
\label{fig:newlhcbqveksmu1l1}
\end{figure}

\begin{figure}[hbp]
\centering
\setlength{\epsfxsize=10cm}
\centerline{\epsffile{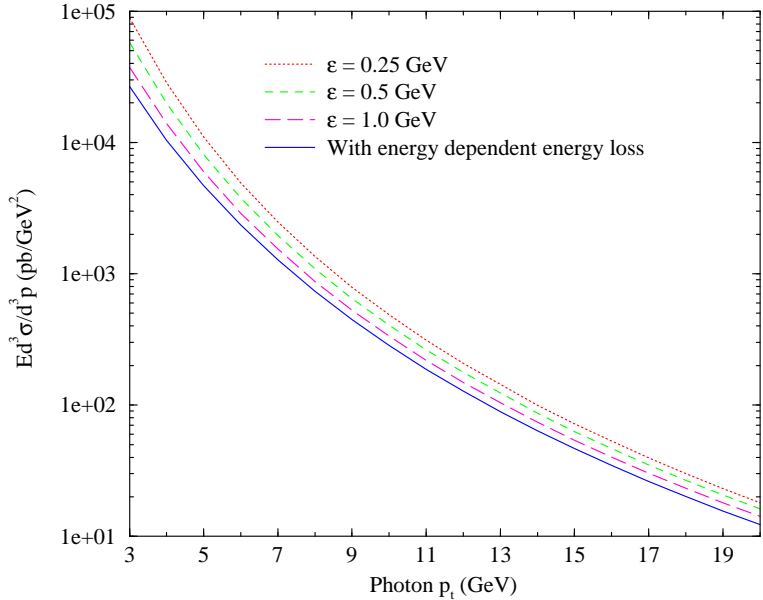}}
\caption{Prompt photon cross section in the central rapidity region 
at $\sqrt{s}=5.5$ TeV obtained with EKS98 shadowing \cite{eks98} 
and with different 
values for parton energy loss .}
\label{fig:newlhc21mu1l1closs}
\end{figure}

We show our results for the prompt photon cross section in nuclear 
collisions at LHC energies 
in Figures (\ref{fig:newlhcbqvmu1l1}), (\ref{fig:newlhcbqveksmu1l1}) and 
(\ref{fig:newlhc21mu1l1closs}).  We use different forms of nuclear shadowing
and energy-dependent energy loss with 
$\mu^2=1$GeV$^2$ and $\lambda_a=1$fm (Figures 5 and 6) or a 
constant energy loss (Figure 7).  
We find nuclear effects at LHC to be 
strikingly large. 
The nuclear cross sections can be reduced by $90\%$  at 
$p_t=3$ GeV and $50\%$ at $p_t=20$ GeV for the energy dependent energy loss
scenario. The constant energy loss of $2$ GeV per scattering leads to
the similar size effects while smaller energy losses per scattering, such 
as $1$ GeV, 
$0.5$ GeV and $0.25$ GeV lead to smaller suppression of the nuclear cross
section. In Figure (\ref{fig:newlhcbqvmu1l1})
BQV parameterization of shadowing is used which is $Q^2$ independent.
This reflects itself in almost constant ($\sim 30\%$) contribution 
of shadowing to the suppression of the nuclear cross section at all $p_t$.
On the other hand, energy loss effects become smaller at larger $p_t$'s 
($\sim 50\%$ at $p_t=3$ GeV and $\sim 20\%$ at $p_t=20$ GeV).  

In Figure (\ref{fig:newlhcbqveksmu1l1}) we use both the BQV \cite{bqv} 
and the EKS98 \cite{eks98} 
parameterization of shadowing. The EKS98 parameterization has 
$Q^2$ 
dependence and distinguishes between quark and gluon shadowing while
the BQV parameterization is $Q^2$ independent and does not distinguish
between quark and gluon shadowing. The difference at low $p_t$ 
($p_t \sim 3$ GeV) 
is small, about $4\%$ while at $p_t=20$ GeV it is about $20\%$.
This is mainly due to the $Q^2$ dependence of EKS98 parameterization.

In 
Figure (\ref{fig:newlhc21mu1l1closs}) we show 
the dependence of the photon cross section 
on the form of energy
loss used, by considering 
different values
of the constant energy loss.  
For comparison, we also show the energy-dependent energy loss case with
$\mu^2 =1 GeV^2$ and $\lambda_g = \lambda_q =1 $fm. It is clear that
the magnitude of the nuclear cross section is sensitive to the magnitude
of the energy loss.

\begin{figure}[htp]
\centerline{
    \epsfxsize 3.0 truein \epsfbox{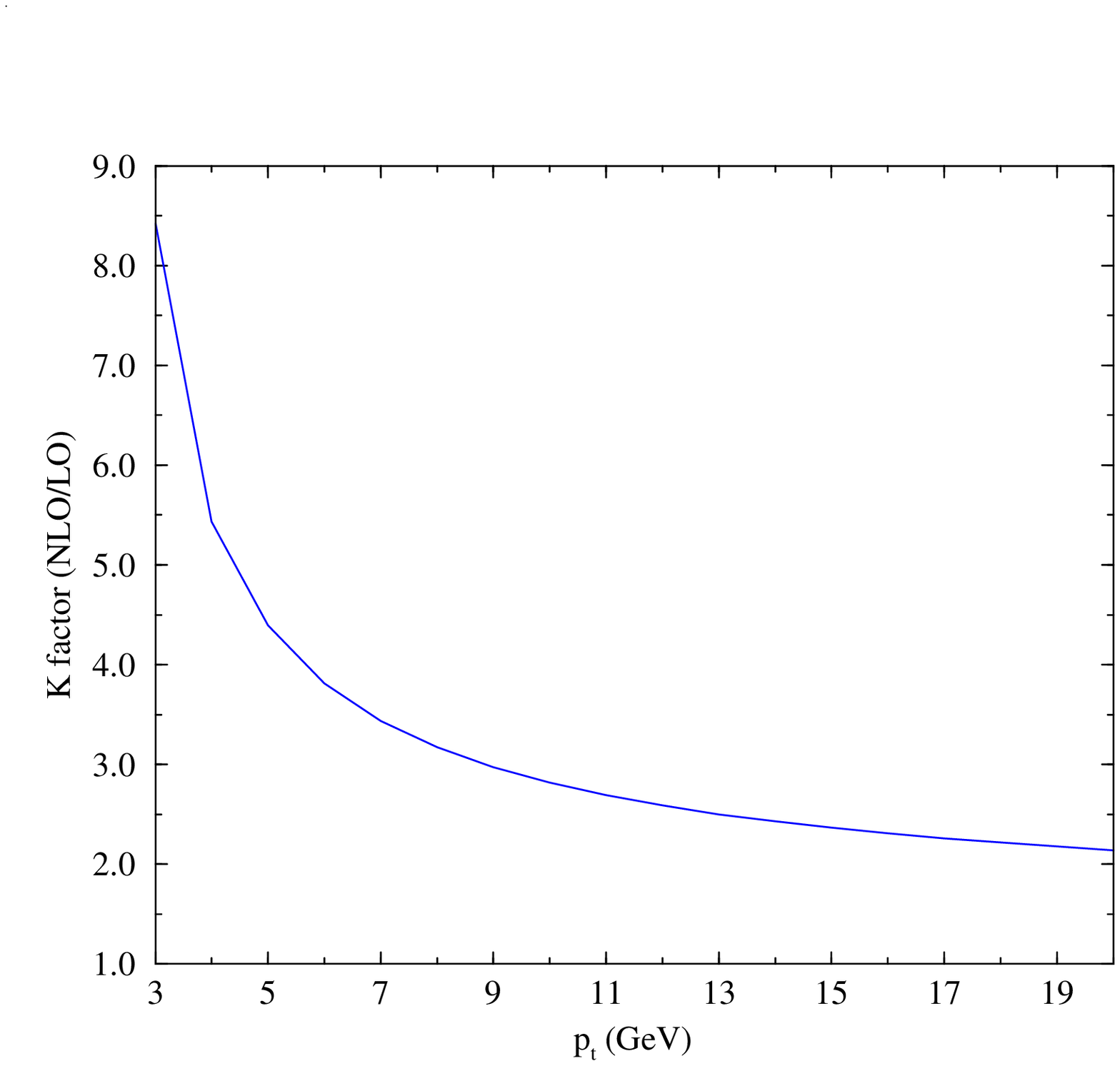}
    \hskip 0.1 truein
    \epsfxsize 3.0 truein \epsfbox{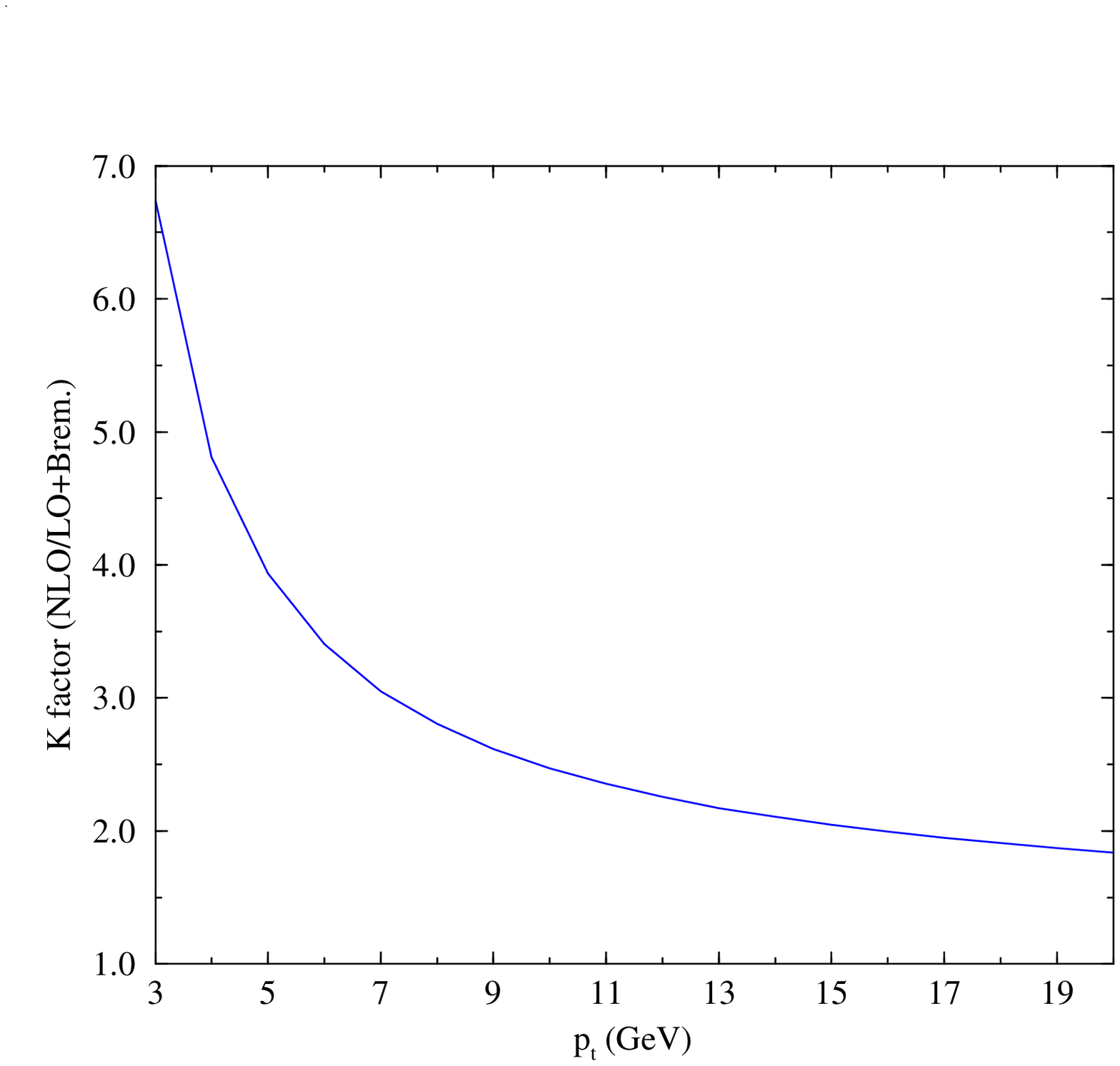}
 }
\caption{a) The nuclear K-factor defined as the ratio of NLO
prompt photon cross section in heavy-ion collisions,
$E{d\sigma \over d^3p}$, to the LO cross section
and b) the K-factor defined as the ratio of NLO to the LO plus
bremsstrahlung cross sections. 
Parton energy loss is
taken to be energy dependent,
$\epsilon^j_a = \alpha_s \sqrt{\mu^2\lambda_a E^j_a}$,
with
$\mu^2=1$GeV$^2$ and $\lambda_a=1$fm.
}
\label{fig:newlhckfnlo}
\end{figure}

\begin{figure}[htp]
\centering
\setlength{\epsfxsize=8cm}
\centerline{\epsffile{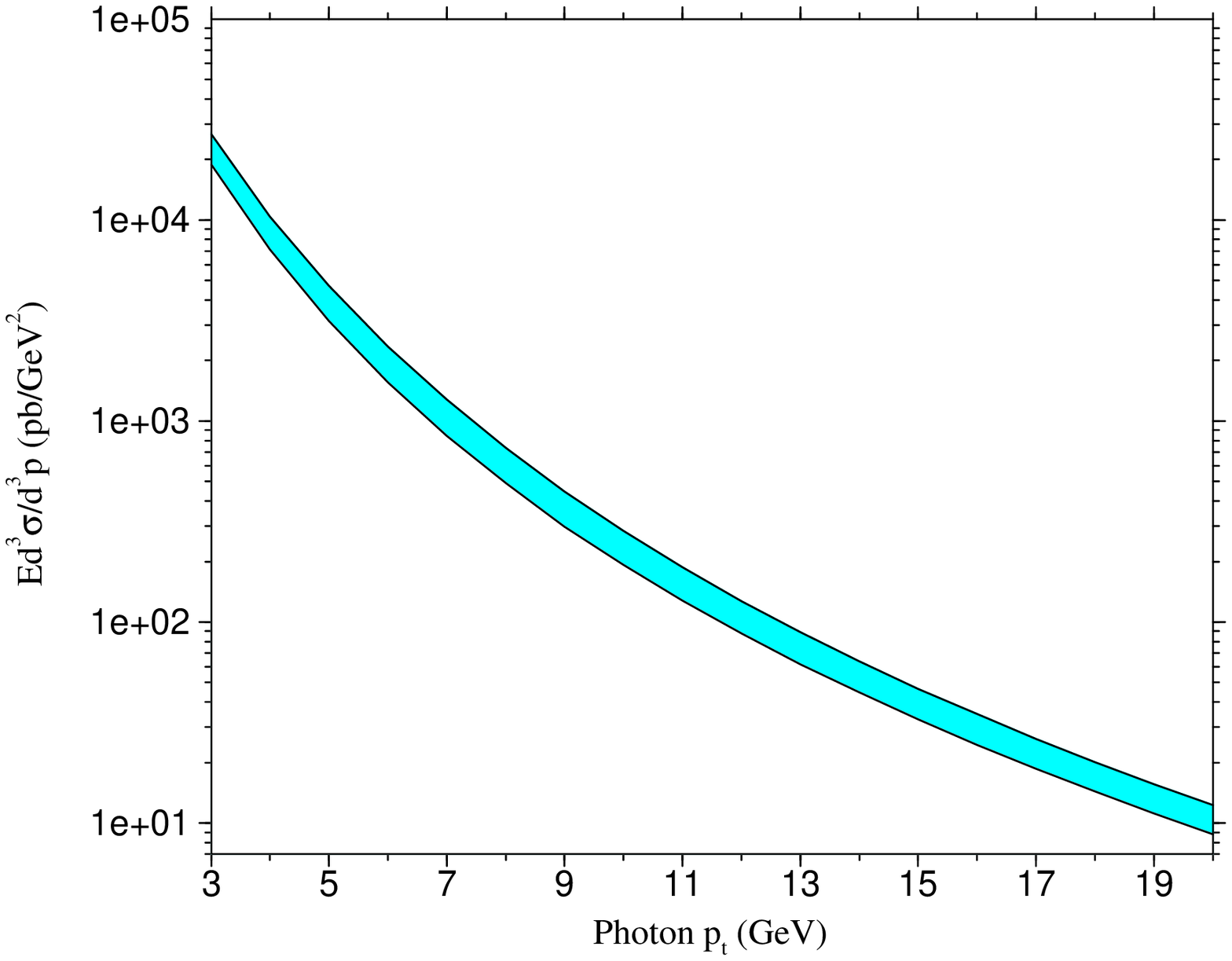}}
\caption{The uncertainty of the photon cross section due to 
varying factorization, renormalization and 
fragmentation scales ($Q=M=Q_F$) 
between $0.5p_t$ and $2p_t$.  
Parton energy loss is
taken to be energy dependent, 
with
$\mu^2=1$GeV$^2$ and $\lambda_a=1$fm.
}

\label{fig:newlhc21scaledep}
\end{figure}

To illustrate the importance of doing the full next-to-leading order 
calculation of prompt photon production at the LHC, we show 
the nuclear $K$-factors, defined as the ratio of $NLO/LO$ and as the 
ratio of $NLO/(LO+Brems.)$ cross sections in Figure (\ref{fig:newlhckfnlo}).
It should be clear that the higher order corrections (beyond NLO) 
are important at low $p_t$ and that our analysis is only qualitatively
reliable there. The dependence of cross section on the choice of renormalization, 
factorization and fragmentation scales is shown 
in Figure (\ref{fig:newlhc21scaledep}).  Varying scales from
$0.5 p_t$ to $2.0 p_t$ results in theoretical uncertainty of 
about 30\%.  
This indicates the importance of including higher order terms, as does the 
large $K$ factor values presented 
in Figure (\ref{fig:newlhckfnlo}).  We 
refer the reader to 
\cite{aur} for a complete analysis of scale dependence of 
prompt photon cross section in $pp$ collisions. 
 
\begin{figure}[htp]
\centering
\setlength{\epsfxsize=8cm}
\centerline{\epsffile{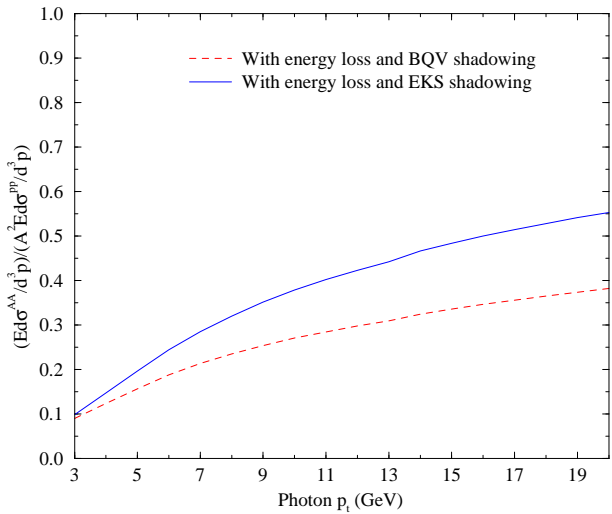}}
\caption{The rescaled prompt photon cross section in the central rapidity 
region at 
$\sqrt{s}=5.5$ TeV using BQV \cite{bqv} and EKS98 \cite{eks98} shadowing and 
energy-dependent energy loss with 
$\mu^2=1$GeV$^2$ and $\lambda_a=1$fm.}
\label{fig:newrslhcbqveks}
\end{figure}

We find that nuclear effects at LHC are significant and
should be easily detectable. Nuclear shadowing effects are 
large and need to be better understood. In Figure 
(\ref{fig:newrslhcbqveks}) we show the rescaled nuclear cross section
using the BQV and EKS98 nuclear shadowing. The difference between
the two parameterizations of nuclear shadowing is significant 
at large $p_t$.
This raises the possibility that one could study the
$Q^2$ dependence of nuclear shadowing by measuring the $p_t$ spectrum
of prompt photons since this difference is independent of the form 
of the energy loss per scattering used.

Current parameterizations
of nuclear shadowing in the kinematic region appropriate to LHC  energies 
are
extrapolations from low energy fixed target data and subject to large
uncertainties. A precise quantitative knowledge of the nuclear structure
functions at the small $x$, large $Q^2$ kinematic region is crucial. A
lepton-nucleus collider such as eRHIC is urgently needed.

\begin{figure}[htp]
\centering
\setlength{\epsfxsize=8cm}
\centerline{\epsffile{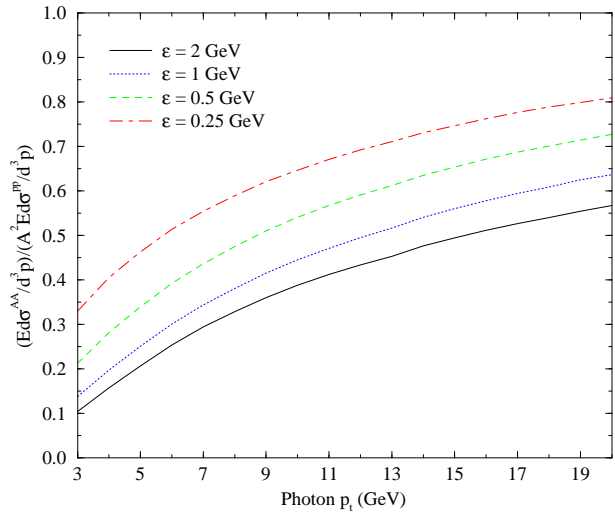}}
\caption{The rescaled prompt photon cross section in the central rapidity 
region at 
$\sqrt{s}=5.5$ TeV obtained using 
constant energy loss and EKS98 \cite{eks98} 
nuclear shadowing form.} 
\label{fig:newrslhcelossvspt}
\end{figure}

Energy loss effects are also large. In Figure  (\ref{fig:newrslhcelossvspt})
we show the rescaled nuclear cross section for the case of constant
energy loss.  
Whether one can distinguish
experimentally between different energy loss scenarios  will depend on
our precise knowledge of nuclear structure functions and also on
the precision of the prompt photon measurements at RHIC and LHC.
Distinguishing prompt photons from those coming from decays of pions
and eta's is notoriously difficult.  Also, one expects that the 
calculation of the ratio of prompt photons to pions would reduce some of 
theoretical uncertainties such as scale dependence and  
intrinsic $k_t$ effects, thus making the NLO calculation even more 
reliable \cite{mike}. We intend to calculate this ratio in the near 
future \cite{jos2}.

\vskip 0.1true in

\leftline{\bf Acknowledgments} 

We would like to thank P. Aurenche and M. Werlen for providing us with 
the fortran routines for calculating double differential
distributions for photon production in hadronic collisions and for 
many useful discussions. J. J-M would like to thank S. Jeon, M. Tannenbaum
and X-N. Wang for many helpful discussions. J. J-M is grateful to LBNL 
nuclear theory group for the use of their computing resources.
This work was supported in part through 
U.S. Department of Energy Grants Nos. DE-FG03-93ER40792 and 
DE-FG02-95ER40906.  
     
\leftline{\bf References}

\renewenvironment{thebibliography}[1]
        {\begin{list}{[$\,$\arabic{enumi}$\,$]}  
        {\usecounter{enumi}\setlength{\parsep}{0pt}
         \setlength{\itemsep}{0pt}  \renewcommand{\baselinestretch}{1.2}
         \settowidth
        {\labelwidth}{#1 ~ ~}\sloppy}}{\end{list}}

\end{document}